\documentclass[sigconf]{acmart}

\AtBeginDocument{%
  }

\setcopyright{acmcopyright}
\copyrightyear{2023} 
\acmYear{2023} 
\setcopyright{acmlicensed}\acmConference[SVR '23]{Symposium on Virtual and Augmented Reality}{November 6--9, 2023}{Rio Grande, Brazil}
\acmBooktitle{Symposium on Virtual and Augmented Reality (SVR '23), November 6--9, 2023, Rio Grande, Brazil}
\acmPrice{15.00}
\acmDOI{10.1145/3625008.3625051}
\acmISBN{979-8-4007-0943-2/23/11}

\usepackage{gensymb}
\usepackage{enumitem}

\begin{document}

\title{Real-Time Viewport-Aware Optical Flow Estimation in 360-degree Videos for Visually-Induced Motion Sickness Mitigation}

\author{Zekun Cao}
\email{zacherycao@gmail.com}
\affiliation{%
  \institution{Dept. of Mechanical Engineering and Materials Science\\Duke University}
  \city{Durham}
  \state{NC}
  \country{USA}
}

\author{Regis Kopper}
\email{kopper@uncg.edu}
\orcid{0000-0003-2081-7061}
\affiliation{%
  \institution{Dept. of Computer Science\\UNC Greensboro}
  \city{Greensboro}
  \state{NC}
  \country{USA}
}

\begin{abstract}
Visually-induced motion sickness (VIMS), a side effect of perceived motion caused by visual stimulation, is a major obstacle to the widespread use of Virtual Reality (VR). Along with scene object information, visual stimulation can be primarily indicated by optical flow, which characterizes the motion pattern, such as the intensity and direction of the moving image. We estimated the real time optical flow in 360-degree videos targeted at immersive user interactive visualization based on the user's current viewport. The proposed method allows the estimation of customized visual flow for each experience of dynamic 360-degree videos and is an improvement over previous methods that consider a single optical flow value for the entire equirectangular frame. We applied our method to modulate the opacity of granulated rest frames (GRFs), a technique consisting of visual noise-like randomly distributed visual references that are stable to the user's body during immersive pre-recorded 360-degree video experience. We report the results of a pilot one-session between-subject study with 18 participants, where users watched a 2-minute high-intensity 360-degree video. The results show that our proposed method successfully estimates optical flow, with pilot data showing that GRFs combined with real-time optical flow estimation may improve user comfort when watching 360-degree videos. However, more data are needed for statistically significant results.
\end{abstract}



\maketitle

\section{Introduction}
Immersive display systems such as Head-Mounted Displays (HMDs) and the increasing accessibility of omnidirectional cameras have paved the way for a surge in the prevalence of 360-degree videos. These videos, offering a view that completely surrounds the user, represent a significant and growing portion of virtual reality (VR) applications worldwide. Applications in education~\cite{shadiev2022review}, training~\cite{patel2020developing}, and live entertainment~\cite{shafi2020360} attest to the popularity and potential utility of 360-degree video experiences. However, such experiences are subject to visually-induced motion sickness (VIMS)~\cite{7996611, rupp2019investigating}.

Similar to traditional motion sickness, VIMS presents comparable symptoms but results from visual stimuli rather than motion. Supported by the Sensory Conflict Theory, VIMS happens when inconsistencies occur between simultaneous sensory stimuli. Under typical daily circumstances, humans primarily perceive motion through a harmonious interplay of visual and vestibular sensory inputs, enabling smooth cognitive processing~\cite{peterka2018sensory}. In this context, the vestibular system detects linear and rotational motion changes, whereas the visual system identifies movements by recognizing changes in the images perceived by the eyes.

However, in an immersive simulated environment such as a 360-degree video, this alignment can break down. While our vestibular system suggests that we remain stationary, the eyes are exposed to dynamic and potentially fast-moving visual stimuli, indicating motion. This sensory conflict challenges our cognitive processing capabilities and ultimately leads to discomfort or motion sickness~\cite{keshavarz2014visually}. The immersive nature and sensory complexity of 360-degree videos, although providing a unique user experience, makes them particularly prone to inducing VIMS.

A leading trigger for VIMS is optical flow~\cite{doi:10.1080/10407413.2014.958029}--the apparent motion of brightness patterns in an image sequence. In real-time interactive VR, the scene is composed of 3D computer graphics elements, and it is trivial to estimate the optical flow experienced by the user by tracking the movement of the virtual camera and other objects in the virtual environment. However, in the case of 360-degree video, where the content is pre-recorded and only the orientation of the user viewpoint can be controlled, estimating the experienced optical flow becomes more challenging.  In this paper, we propose a method that includes a pre-processing step and a real-time component for the estimation of the optical flow experienced by a user in VR during the experience of a 360-degree video.

The main contribution of this study is the development of a new method to estimate the optical flow experienced by a user in VR during the viewing of a 360-degree video. During the experience of a 360-degree video, real-time optical flow estimation offers the unique opportunity to mitigate VIMS when it is most likely to occur. We tested this by conducting a short, single-session pilot study, and found non-significant improvements in mitigating VIMS symptoms using our proposed method.

\section{Related Work}
In this section, we cover the literature relevant to the association between optical flow and VIMS, existing methods to determine optical flow in 360-degree videos, and methods used to mitigate VIMS.

\subsection{Optical Flow and VIMS}
Optical flow is a fundamental concept in computer vision and perception that refers to the apparent motion of objects in a sequence of images~\cite{horn1981determining}. It characterizes motion patterns, including the intensity and direction of moving images. In the context of VR, optical flow becomes particularly important, as it can indicate the motion visually experienced by the user and the respective sensory conflict that may arise when high optical flow is coupled with stationary physical motion. Factors that influence optical flow include details of object motion, amount of the viewport occupied by moving objects, and self-motion of the user~\cite{beauchemin1995computation}. 

Previous research has found that stationary visual scenes cause lower VIMS severity than moving scenes~\cite{lo2001cybersickness}. With this information, optical flow becomes a potential predictor for VIMS in 360-degree videos~\cite{padmanaban2018towards, lee2019motion}. For example, high-intensity self-motion in a virtual environment typically generates high volumes of optical flow, whereas a small object with high-intensity motion does not. The former condition is more provocative than the latter. Inspired by this observation, Park \textit{et al.}~\cite{park2022mixing} nullified the optical flow exposed to users by visually mixing artificial optical flow directed reverse to the virtual visual motion. They found a significant reduction in VIMS by reducing the optical flow using their proposed method.

\subsection{Optical Flow and 360-degree videos}
The estimation of optical flow with 360-degree videos is more challenging than that with interactive VR scenes. As there are no computer graphics behind the generation of video images, optical flow estimation must rely solely on the visual information in the video itself.

Prior methods either used the optical flow of the entire equirectangular (unwrapped) video frame as the input image~\cite{padmanaban2018towards, park2022mixing} or captured eye-tracking and head-tracking data to calculate the optical flow in post-processing~\cite {lee2019motion}. There are significant drawbacks to these methods. 

Using the optical flow of the entire equirectangular frame as the input image~\cite{padmanaban2018towards, park2022mixing} may lead to inaccurate estimates for two reasons. First, the equirectangular frame is largely distorted in the poles (see \autoref{fig:equireactangular}), which causes perceived motion (and, by extension, optical flow) to be much higher in the peripheral regions than in the central regions of the equirectangular footage. Second, the unwrapped frame does not consider the user's current viewport, which varies during a 360-degree video experience and represents only a small portion of the video frame.

Analyzing the optical flow as a post-processing step of a VR experience~\cite{lee2019motion} can lead to important insights into the relationship between optical flow and simulator sickness; however, it does not help mitigate VIMS effects in real time.

\subsection{VIMS Mitigation}
There is consistent evidence that higher levels of field of view (FOV) cause more VIMS~\cite{keshavarz2014visually}. Considering this, Fernandes and Feiner~\cite{fernandes2016combating} proposed a method, called tunneling, to reduce VIMS by manipulating the FOV experienced by the user. Using the virtual motion and rotation speed in an interactive virtual environment as a proxy for optical flow, the tunneling adds software-based peripheral blinders to effectively reduce the FOV of the scene whenever provocative motion is detected. The method showed a statistically significant reduction in reported VIMS. Tunneling has become popular across VR content creation; however, it cannot be directly applied to 360-degree videos, because the user does not control virtual motion or rotation speed in such experiences.

In a recent study on the interplay between user perception and immersive video images, Islam \textit{et al.} proposed a novel deep fusion network to predict VIMS based on 360-degree videos, eye-tracking, and head-tracking data~\cite{islam2021cybersickness}. The study showed the best prediction accuracy when eye-tracking and head-tracking data were used and confirmed that VIMS was highly correlated with the image perceived by the users. However, this approach does not directly estimate the optical flow experienced by a user during a 360-degree video VR experience. To address this limitation, we propose a method to estimate the viewport-based instantaneous optical flow perceived by a user during the VR experience of a 360-degree video.

Rest frames~\cite{prothero1998role} are visual references in a virtual environment that remain static with respect to the real world. A cockpit is a common example of a rest frame used in VR applications. Rest frames have been found to reduce VIMS by providing a stable visual reference point for the user~\cite{8446210}. An alternative to rest frames is granulated rest frames (GRFs)~\cite{cao2021granulated} which are visual noise-like randomly distributed visual elements that can be dynamically displayed to reduce VIMS. GRFs have been shown to be more effective in search tasks than reduced FOV~\cite{cao2021granulated} and may combine the benefits of reduced FOV and rest frames in lowering VIMS, although there is no statistical evidence showing the effectiveness of GRFs in mitigating VIMS.

Based on these findings, this study proposes a novel method to estimate the optical flow experienced by a user in VR during the experience of a 360-degree video. The proposed method improves upon previous approaches by considering the user's current viewport and estimating the instantaneous visual flow during 360-degree video experiences.

\section{Real Time Estimation of Optical Flow}
In this section, we detail the steps necessary for the accurate estimation of optical flow in 360-degree videos.

\subsection{Rationale}
Understanding the optical flow a user experiences while immersively watching a 360-degree video can potentially inform the system about the likelihood that a user may be at the onset of experiencing VIMS. However it is not trivial to compute the optical flow of videos of this nature owing to the unbounded viewport orientation and the unique challenges posed by the spherical projection and the distortion present in unwrapped 360-degree video frames. 

When watching a 360-degree video in an HMD, the user has the freedom to look in any direction, and the content they experience can be different every time they watch the video again. For example, let us consider a user watching a 360-degree video of a crowded outdoor area. The user may choose to focus their attention on different individuals or objects within the scene at different times. Each time the user watches the same video, they may choose to focus on different areas of interest, resulting in different patterns of motion in their field of view. This, in turn, causes a unique optical flow pattern each time a video is watched. Certainly, we do not want a single optical flow value for each unwrapped frame of the entire 360-degree video. In an extreme example, if the user stares at the ground, the optical flow could be zero, whereas if they look at a large train moving by at the same moment, the optical flow could be significant and dynamic. Thus, it is necessary to develop a method that can estimate the optical flow that considers the instantaneous viewport a user has at each moment during the experience of a 360-degree video.

\begin{figure}[!bht]
    \centering 
    \includegraphics[width=1\columnwidth]{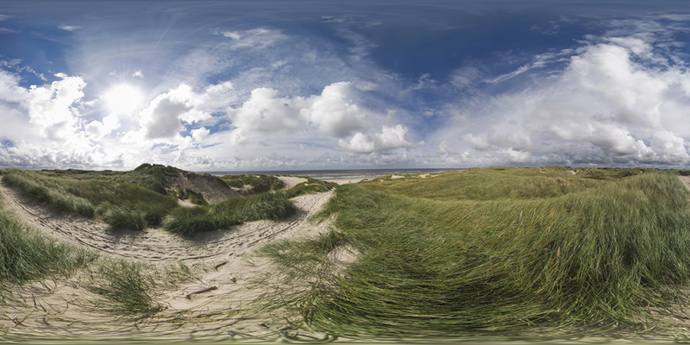}
    \caption{Example of an equirectangular image of a grassland. Image distortions are apparent along the latitudes.}
    \label{fig:equireactangular}
\end{figure}

Equirectangular projection, the most common format for representing 360-degree images, suffers from severe distortions along latitudes when mapping a spherical image onto a plane (see \autoref{fig:equireactangular}). However, users typically experience 360-degree images immersively with an HMD. In other words, the closer an image from the user's perspective is to the polar regions, the more pixels will represent that viewport in the equirectangular frame. Such distortion causes the areas around the poles of the equirectangular image to be stretched and skewed, leading to inaccuracies in optical flow estimation if it is not accounted for.

To address these challenges, our proposed method involves two high-level steps. The first is to segment the 360-degree video into multiple sliding windows, generating viewport-sized tiles that overlap, ensuring that the four nearest sliding windows have large amounts of overlap with any possible viewport. The second step is to account for the distortions present in the equirectangular projection to calculate the optical flow. For this, we propose applying a series of transformations to correct the radial distortion on 360-degree videos.

\subsection{Optical-Flow Calculation}

A 360$^\circ$ video equirectangular projection maps 360$^\circ$ horizontally and 180$^\circ$ at the vertical dimension. To obtain the real optical flow users are exposed to when watching the video, the first thing that needs to happen is to split and convert the equirectangular image into pieces the size headset's field of view--the most extensive area the headset can cover in the equirectangular image, also known as the user's viewport.

Here, we used a sliding window to scan the video by 15$^\circ$ horizontally and 7.5$^\circ$ vertically per step in the pre-calculation process. The sliding window represented the viewport of the HMD and matched its FOV. The local coordinate system of the sliding window was placed at the center of the represented viewport. For example, the final converted video for the sliding window with center at $(\phi, \theta)$ is the view the user perceives in the HMD when their head has pitch $\phi$ and yaw $\theta$. The same calculation process was performed for each frame of the video.

Below are the steps to convert the image in each sliding window:
\begin{enumerate}[leftmargin=*,label=\arabic*.]
    \item Get the output image using center $(c_x,\ c_y)$ :
    \begin{equation}\label{eq:3}
        c_x = \frac{W}{2}
    \end{equation}
    \begin{equation}\label{eq:4}
        c_y = \frac{H}{2}
    \end{equation}
where the left eye’s image in the HMD has $W \times H$ resolution.
    \item Calculate the horizontal FOV  ($w_{FOV}$) and the vertical FOV ($h_{FOV}$) in the output image:
    \begin{equation}\label{eq:5}
        w_{FOV} = FOV
    \end{equation}
    \begin{equation}\label{eq:6}
        h_{FOV} = FOV\times\frac{H}{W}
    \end{equation}
    where $FOV$ is the headset’s left eye's horizontal FOV.
    \item Get the units of the view represented by each horizontal pixel at perspective image  ($w_{Ratio}$):
    \begin{equation} \label{eq:7}      w_{Ratio} = 2 \times \frac{\tan{\frac{w_{FOV}}{2}}}{W}.
    \end{equation}
    
    \item Determine the units of the view represented by each vertical pixel at perspective image ($h_{Ratio}$):
    \begin{equation}\label{eq:8}
    h_{Ratio} = 2 \times\frac{ \tan{\frac{h_{FOV}}{2}} }{H}.
     \end{equation}
     
    \item Convert the pixel with coordinate $\Vec{v} = (x,\ y,\ z)$ in the converted sub-image to position $\Vec{v'}=(x',\ y',\ z')$ in the equirectangular image. Here $\Vec{v}$ and $\Vec{v'}$ represent the pixel’s coordinates before and after the conversion, respectively.
    \begin{equation}\label{eq:9}
    \Vec{v'} = \frac{((x-c_x) \times w_{Ratio}, (y-c_y) \times h_{Ratio}, 1)}{D}
    \end{equation}
    where
    \begin{equation}\label{eq:10}
    D =  \sqrt{1^2 + [(x-c_x) \times w_{Ratio}]^2 + [(y-c_y) \times h_{Ratio}]^2}
    \end{equation}
    is the total distance between the camera and the pixel.
    \item Apply Rodrigues' rotation formula to the coordinate as if the camera was rotated by $(\phi\text{(pitch)},\theta\text{(yaw)}))$.
    \begin{equation}\label{eq:11}
    \begin{split}
        \vec{v''} = \vec{v'}\cos{\phi} + (\vec{k_x}\times \vec{v'})\sin{\phi}+ \vec{k_x}(\vec{k_x}\cdot\vec{v'})(1-\cos{\phi})
    \end{split}
    \end{equation}
    \begin{equation}\label{eq:12}
    \begin{split}
        \vec{v'''} = \vec{v''}\cos{\theta} + (\vec{k_y}\times \vec{v''})\sin{\theta}+ \vec{k_y}(\vec{k_y}\cdot\vec{v''})(1-\cos{\theta})
    \end{split}
    \end{equation}
    where: $\vec{v''}$ is $\vec{v'}$ after pitch rotation. $\vec{v'''}$ is $\vec{v''}$ after yaw rotation. $\vec{k_x}$ and $\vec{k_y}$ are the x-axis and y-axis.
    
    \item Transform the cartesian coordinates from  \autoref{eq:11} back to spherical coordinates, which are the longitude and latitude of the pixel, as follows:
    \begin{equation}\label{eq:13}
        \text{Latitude} = \arcsin{\vec{v'''}_y}
    \end{equation}
    \begin{equation}\label{eq:14}
        \text{Longitude} = \arctan{\frac{\vec{v'''}_x}{vec{v'''}_z}}
    \end{equation}
    \item Use the latitude and longitude from \autoref{eq:13} and \autoref{eq:14} to calculate the position of the pixel in the equirectangular image, as follows:
    \begin{equation}\label{eq:15}
        \text{x}_{\text{new}} = \text{Longitude}\times \frac{W_{equirectangular}}{2}  + \frac{W_{equirectangular}}{2}
    \end{equation}
    \begin{equation}\label{eq:16}
        \text{y}_{\text{new}} = \text{Latitude}\times \frac{H_{equirectangular}}{2} +\frac{H_{equirectangular}}{2}
    \end{equation}
    \item Use 4$\times$4 neighboring pixels of the image around position $(\text{x}_{\text{new}},\ \text{y}_{\text{new}})$ to interpolate the required pixel.
\end{enumerate}

\subsection{Optical Flow Prediction}
Once the sliding window videos were converted, each frame of each sliding window was calculated using a pre-trained deep learning model to predict its optical flow. Because the focus of this research was not to create a new optical flow prediction algorithm, we used FlowNet 2.0~\cite{ilg2017flownet} to calculate the optical flow of the 360-degree videos we used in our testing. The model is based on end-to-end optical flow estimation with convolutional neural networks (CNNs) and has better performance and efficiency than FlowNet~\cite{dosovitskiy2015flownet}.

\subsection{Computational Cost} \label{sec:compcost}
We used the Google Colab Pro service, which had Nvidia Tesla P100-PCIE-16GB GPUs, Intel(R) Xeon(R) CPUs @ 2.20GHz, and 24 GB RAM to pre-process the optical flow estimation. The pre-calculation computational cost was approximately 6 s for each viewport within each video frame for each GPU. This computational cost is high, with the computation performed over time using multiple GPUs. As this was a pre-processing step, such performance issues were not a concern.

\subsection{Real-Time Optical Flow Estimation}\label{subsubsec: OFEst}
\begin{figure*}[!ht]
    \centering 
    \includegraphics[width=1\textwidth]{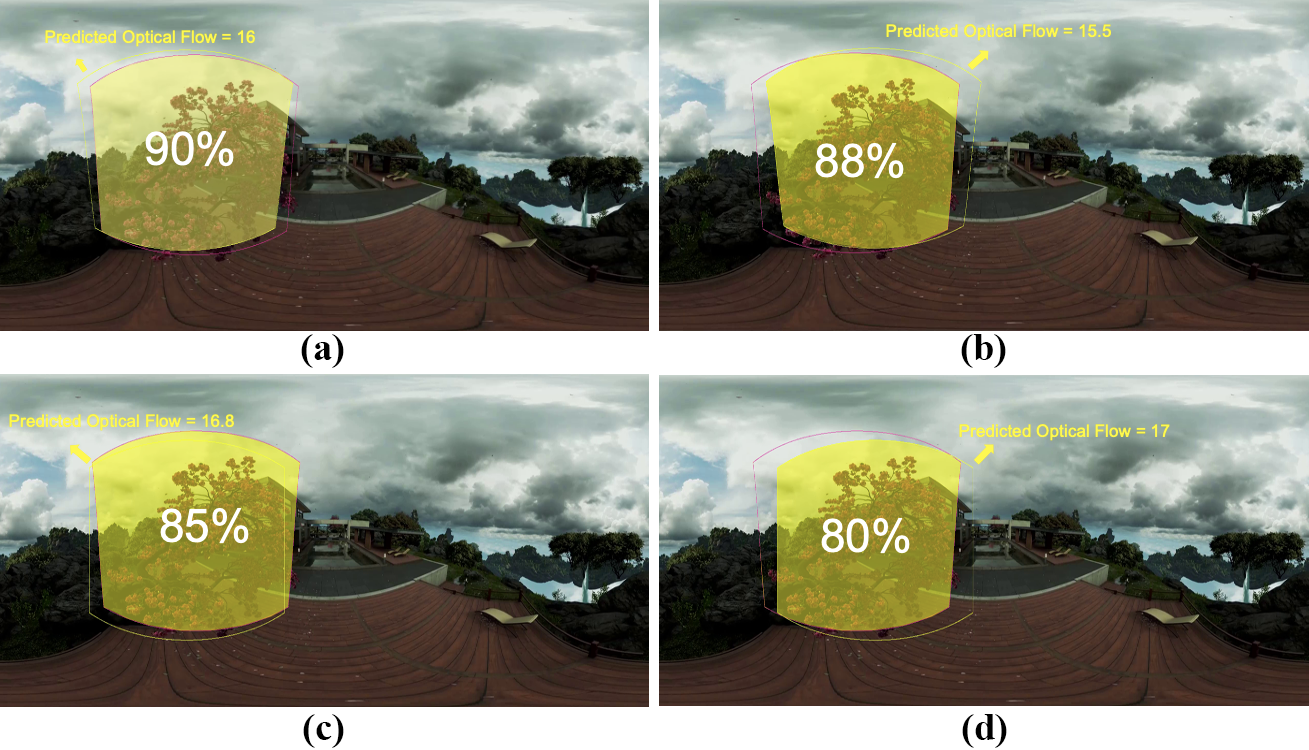}
    \caption[Predicted optical flow of user's viewport.]{Predicted optical flow of the user’s viewport (pink circles). Images (a) to
(d) are four different sliding windows (yellow circles) overlapping the current viewport (yellow-shaded areas).}
    \label{fig:viewport}
\end{figure*}

While experiencing 360-degree videos, users can look in any direction, and the discrete nature of sliding windows does not capture all the possible view perspectives. Therefore, to approach the exact predicted optical flow of any viewport, the real-time optical flow was estimated using the weighted average of all sliding windows overlapping with the user's current viewport. The weight of each sliding window was calculated as the percentage by which each viewport overlapped with the current sliding window. For example, in \autoref{fig:viewport} the viewport, shown in pink, has four overlapping sliding windows (highlighted in yellow). They have optical flow values of 16, 15.5, 16.8, and 17, respectively. The overlap of each sliding window is 90\%, 88\%, 85\%, and 80\%, respectively. Thus, the estimated perceived optical flow (EPOF) in the example is calculated as
\begin{align*}
EPOF   & =\ \frac{\sum\limits_{n=1}^{k} POF_n \times OL_n}{\sum\limits_{n=1}^{k} OL_n} \\
       & =\ \frac{90\% \times 16 + 88\% \times 15.5+ 85\% \times 16.8+ 80\% \times 17}{90\% + 88\% + 85\% + 80\%} \\
       & \approx\ 16.3.
\end{align*}

\section{Pilot Study: VIMS Mitigation}
The proposed method allows us to obtain an accurate estimate of the predicted optical flow during the experience of a 360-degree video in VR in real time. To test the validity of our method and gain initial evidence of the effectiveness of real-time optical flow adjustment in mitigating visually induced motion sickness, we conducted a small pilot study.

\subsection{360-degree Video Selection and Pre-Processing}
To enable the evaluation of our proposed method, we selected the 360-degree video that participants would experience from a public dataset, shared by Padmanaban \textit{et al.}~\cite{padmanaban2018towards}. This dataset was chosen because it was evaluated in the context of VIMS prediction and optical flow analysis. Padmanaban \textit{et al.} used a machine learning model to predict VIMS based on optical flow and other parameters. However, their approach only considered the entire equirectangular frame for optical flow prediction, which does not account for viewport or distortions in the source frame.

Our final selection was a 2-minute stereoscopic 360-degree video that appended the ``Cartoon Coaster'' and ``Ship'' videos from the dataset. These videos had moderate-to-high VIMS ground truths as measured in a user study~\cite{padmanaban2018towards}. The combined 2-minute video was processed in Python with the Google Colab Pro service (see \autoref{sec:compcost}) over several days. The resulting sliding window optical flow matrix was exported as an array that was used in Unity for real-time optical flow estimation (\autoref{subsubsec: OFEst}).

\subsection{Apparatus}

\subsubsection{Hardware}

An HTC Vive Pro Eye with six degrees-of-freedom (DoF) tracking was the HMD used in our experiments. Because the experience of 360-degree videos does not allow locomotion, only rotational DoF (yaw, pitch, roll) tracking data were used to allow the user to choose where to look. The HMD had a horizontal FOV of 107$^\circ$ and a vertical FOV of 107$^\circ$). This resulted in 198 sliding windows per 360-video frame (18 horizontal $\times$ 11 vertical segments). 

The desktop running the application software was an AMD Ryzen 7 2700X Eight-Core Processor (3.80 GHz) with 16GB RAM and an Nvidia GeForce RTX 2060 running Windows 10. Two Vive controllers were used to input the participant’s ID through a UI in the scene at the beginning of each session. The right controller was used to select each digit number ($0-9$) through ray casting. The trigger button of the left controller was used to confirm the ID. Once the ID was input, the video was automatically started. No interaction beyond head rotation was available during the video experience.

\subsubsection{Software and VIMS Mitigation Strategy}

The experimental virtual environment was written in Unity and contained an initial screen, where the participant entered their study ID, and the 2-minute stereoscopic video mentioned above.

To mitigate VIMS, we adopted the same GRF settings as in Cao \textit{et al.}~\cite{cao2021granulated}, which maximizes visual search ability and VIMS reduction without sacrificing presence. It consists of two main parameters: \textbf{Size}, which represents the amount of FOV (in degrees) that each grain (black circle) covers, and \textbf{Density}, which represents the percentage of FOV that is blocked by grains. Based on Cao \textit{et al.}'s work, the \textbf{Size} parameter was set to $1.5^{\circ}$ of the FOV, and the \textbf{Density} parameter was set to 50\%. We also applied a soft-edged circular cutout at the center of the user’s FOV, which contained a linearly-increased opacity from transparent within the $36^{\circ}$ inner FOV (IFOV) to opaque beyond the $80^{\circ}$ outer FOV (OFOV). 

\begin{figure*}[!ht]
    \centering 
    \includegraphics[width=1\textwidth]{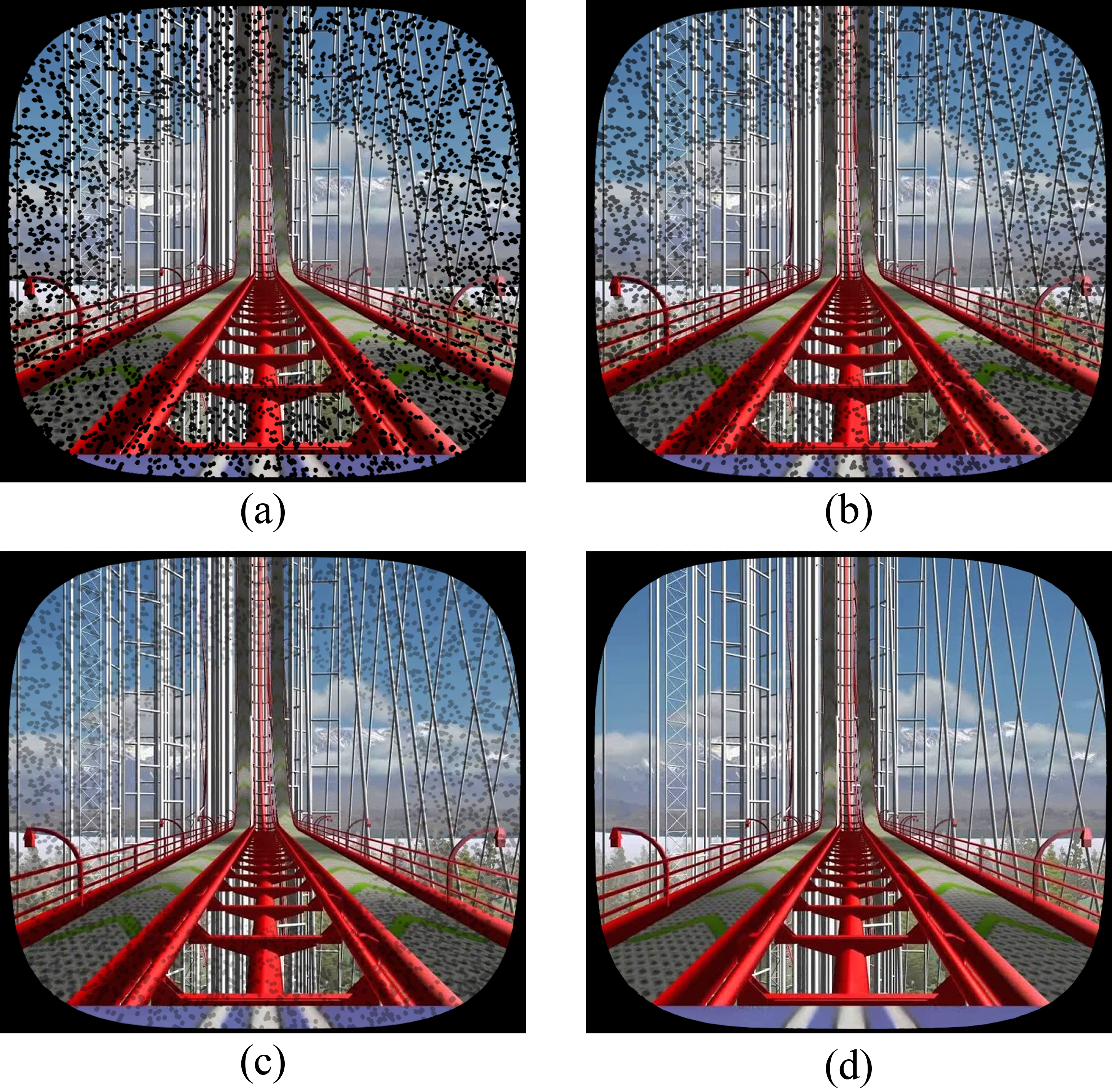}
    \caption{Granulated rest frames with different opacities based on the estimated optical flow of the user's viewport. Images (a) to (d) show GRFs with 100\%, 60\%, 30\%, and 0\% opacity, respectively.}
    \label{fig:GRFopacity}
\end{figure*}

The overall GRFs display opacity was modulated based on the range of optical flow in the entire video. We set the GRFs as fully transparent at the 10\textsuperscript{th} percentile optical flow value and fully opaque at the 90\textsuperscript{th} percentile optical flow value. For an estimated optical flow value between the 10\textsuperscript{th} and 90\textsuperscript{th} percentiles, the GRF opacity was linearly interpolated. With this strategy, the GRF opacity varied continuously based on the optical flow intensity, minimizing interruptions in the user experience. Fig. \ref{fig:GRFopacity} shows one eye's view of different GRF opacities.

\subsubsection{Pilot Study Design}

The experiment followed a one-session between-subjects design with post-study questionnaires as the dependent variables (DVs). The questionnaires used in the study were the Simulator Sickness Questionnaire (SSQ)~\cite{kennedy1993simulator} and the Motion Sickness Susceptibility Questionnaire Short-Form (MSSQ-Short)~\cite{golding2006predicting}.

The only independent variable controlled for in the study was \textbf{Condition} (Granulated Rest Frames--\textit{GR}, no rest frames--\textit{NR}). \textit{GR} participants experienced GRFs in their experimental session, whereas \textit{NR} participants did not. Under COVID-19 University policy, all participants wore face masks during the study.

Each session was allocated 2 min to the immersive 360-degree video experience and 10 min to fill out post-study questionnaires. All participants signed an informed consent form and a background survey to capture demographic and VR experience data before starting their session. The participants were introduced to the environment and learned how to use the devices. At the start of the session, the participants entered their ID using the VIVE controller in their right hand and experienced the 360-degree video, freely looking around by rotating their body and head during the session. 

The study protocol was approved by the Institutional Review Board from the University of North Carolina at Greensboro.

\subsection{Data Collection}

During the study, head rotation data from the HMD were collected at 60 Hz, which coincided with the video's 60 frames per second (fps) update rate. The head rotation data allowed us to calculate the viewport of the user at every frame. Therefore, EPOF was calculated at every frame of the video based on the calculated viewport, as described in \autoref{subsubsec: OFEst}. Because the estimation only required the weighted average across overlapping sliding windows (which existed as an array data structure), there was no impact on the performance of the application, which ran at full frame rate (90 fps).

After finishing the session, participants filled out post-study questionnaires.

\subsection{Participants}

We recruited 18 participants (9 females, 3 non-binaries, ages 18-27, mean$\pm$SD 20.7$\pm$2.88) for the experiment, all of whom finished their assigned session. Nine participants were randomly assigned to one of the \textit{GR} condition or \textit{NR} condition. All the participants were healthy and had normal or corrected-to-normal vision.

\subsection{Results}

Of the 18 participants, three barely moved their heads, which caused the EPOF to be much lower than that of the other participants. Their optical flow range was $[90,650]$, whereas that of the other participants was $[1075,2927]$. We decided to exclude their data from the analysis because unrealistically low optical flow sessions could introduce significant noise to the data because GRFs were triggered by the amount of participants' EPOF. Among the three excluded participants, one was from the \textit{GR} condition.

\subsubsection{Data Transformation}
To account for the uniqueness of each participant's 360-degree video experience, given that they could look in any direction during the session, and to account for varied motion sickness susceptibilities across participants, we transformed the participants' SSQ score using the following equation:
\begin{equation}  \label{eq:transf}
    K_{OF,i} = \frac{K_{i}\times MS_{i}}{Optical\ Flow_{i} \times \sum_{j\in \mathbb{U}_{g}}{MS_{j}}} \times 1000,
\end{equation}
where $K_{OF,i}$ denotes the transformed SSQ total score for participant \textit{i}, $K_i$ represents the original SSQ total score for participant \textit{i}, $MS_{i}$ is participant \textit{i}'s MSSQ-Short score, and $\mathbb{U}_{g}$ represents the set of users assigned to the same group ($g\in \{GR, NR\}$). The same transformation was performed for each subscore of the SSQ, where the nausea subscore was denoted as $N_{OF}$, the oculomotor subscore as $O_{OF}$, and the disorientation subscore as $D_{OF}$. The summation of MS at the denominator in \autoref{eq:transf} was used to normalize the transformed data. By calculating $[K,N,O,D]_{OF}$, more susceptible participants were weighted higher, increasing the sensitivity of our measurement for participants more prone to SSQ symptoms.

\subsubsection{Data Analysis}
Among the 15 data points used in the analysis, we first conducted the Shapiro-Wilk test to determine whether the SSQ score per unit of optical flow was normally distributed. The result showed that it was not normally distributed ( $O_{OF}$: $p = 0.00$, $N_{OF}$: $p = 0.00$, $D_{OF}$: $p = 0.00$, $K_{OF}$: $p = 0.00$). Therefore, we ran the non-parametric Mann-Whitney Test over the results (see \autoref{tab:videoSSQ}). 

\begin{table}[htb]
	\centering
	
	\caption{Post-study SSQ at Study 2.} \label{tab:videoSSQ}
	
	\begin{tabular}{llll}
		\toprule
		\textbf{Subsocre} & \textbf{Means} & \textbf{\textit{p} value}\\
		\midrule
		$N_{OF}$ & & \\
		NR vs. GR & 0.51 vs. 1.04   & $0.12$\\
		$O_{OF}$ & & \\
		NR vs. GR & 1.62 vs. 1.00   & $0.13$\\
		$D_{OF}$ & & \\
		NR vs. GR & 2.42 vs. 1.16   & $0.632$\\
		$K_{OF}$ & & \\
		NR vs. GR & 17.00 vs. 11.99   & $0.39$\\
		\bottomrule
		
	\end{tabular}
\end{table}

There were no main effects between \textit{GR} and \textit{NR} conditions. However, the results showed lower means of Oculomotor, Disorientation, and Total scores for the \textit{GR} condition. Although \textit{NR} condition had a lower Nausea score, Stanney \textit{et al.}~\cite{stanney1997cybersickness} pointed out that VIMS was predominated by Disorientation symptoms and Oculomotor symptoms the least. Thus, the \textit{GR} condition had better performance in the predominant subscore and the total score than the \textit{NR} condition, although the difference was not significant. Additionally, only three participants in the \textit{GR} condition reported that they noticed the GRFs. This is evidence that the implementation of GRFs does not compromise users' feeling of presence. Alongside supporting work on VIMS mitigation through rest frames~\cite{cao2018visually,cao2021granulated}, and through the correlation between optical flow and scene provocativeness, we infer that it is useful to trigger rest frames to alleviate VIMS using real-time optical flow as a proxy.

\section{Discussion}

A central contribution of our work is the potential to generalize VIMS prediction in 360-degree videos by estimating the directly perceived optical flow. We simplified optical flow estimation by leveraging mature computer vision/machine learning models and sliding windows pre-processing. Because the visual stimulus is a direct factor causing VIMS, real-time perceived optical flow can be instrumental in predicting the development of VIMS. Our method allows the optical flow estimation from an approximation of the real-time user's view perspective without distortions, instead of computing the optical flow for the entire equirectangular frame. This allows users to move their heads freely during the experience and avoids the need to freeze head rotation for an accurate calculation of optical flow, as done in prior work~\cite{padmanaban2018towards}.

Our proposed optical flow estimation approach uses pre-calculated and readily available data for fast real-time computation. Previous studies have attempted to predict VIMS using bio-physiological signals or perceived visual content~\cite{padmanaban2018towards,islam2021cybersickness, kim2019deep}. External sensors for bio-physiological data collection often require seated conditions and limited 3D-object manipulation, which can affect the user experience and potentially aggravate VIMS. The perceived visual content is typically collected with a low sampling rate or without considering head orientation data, which does not reflect the real visual content perceived by users during normal use. In contrast, our method uses pre-processed viewport approximations, which allow VIMS estimation based on the actual instantaneous visual flow experienced by the user. Using pre-processed data to predict and alleviate VIMS is highly suitable for standalone HMDs and other VR devices with limited computational resources.

\subsection{Other Applications of Real-Time Optical Flow Estimation}

There are other applications of real-time optical flow estimation beyond the estimation of VIMS in VR experiences. Optical flow estimation has been widely used in various computer vision tasks such as activity recognition and saliency detection. These classical applications of optical flow can be used, for example, for the real-time analysis of 360-degree video, showing users potential areas of interest and directing their attention.

There is also the potential to dynamically distribute GPU resources in rendering frames. For example, an area of the frame with a higher real-time optical flow indicates more complex motion patterns that are typically harder for humans to fully discern. These dynamic regions could be selectively rendered at lower resolutions. At the same time, parts of the frame with more details but lower optical flow indicate smoother motion, which can use more computational resources to achieve higher resolution and offer sharper details.

\subsection{Limitations and Future Work}

Our study offers an initial investigation into the benefit of GRFs in alleviating VIMS. Still, there are limitations that need to be addressed in future studies.

First, participants in the pilot study were limited to watching a short 2-minute 360-degree video, which may not have gone beyond the initial stages of VIMS. This is possibly the reason why we saw no significant differences in the reported sickness between conditions. We were limited by our computational resources, which required several days to process the 2-minute video. Future work should focus on increasing the duration of 360-degree videos and assess whether longer exposure to VR experiences with GRFs leads to significantly less VIMS than those without. Additionally, more efficient methods of optical flow estimation should be sought to allow streamlined generation of real-time optical flow estimates for longer video durations. Faster pre-processing could also permit the use of more narrowly defined sliding window steps to the ones we used for our testing (15$^\circ$ horizontal and 7.5$^\circ$ vertical windows).

In addition to limited exposure time, the sample size of our pilot study was small, which increased noise and limited the generalizability of our findings. Future work should focus on conducting larger studies with longer videos and more participants. Furthermore, the effectiveness of GRFs in reducing VIMS has not yet been fully determined. Using our proposed method with more proven mitigation strategies, such as dynamic FOV modification~\cite{fernandes2016combating} could lead to stronger results. 

In our work, we evaluated the benefit of optical flow-triggered GRFs in mitigating VIMS for passive 360-degree experiences, but not in interactive virtual environments. Future work should focus on leveraging the optical flow in interactive virtual environments to modulate the presentation of GRFs, which could be rendered on a pixel scale. This work would build upon the restricted FOV proposed by Fernandes and Feiner~\cite{fernandes2016combating} and the static rest frames introduced by Cao \textit{et al.}~\cite{8446210}. Both techniques modified the visual stimuli perceived by users; the former limited the amount of perceived visual stimuli by blacking out the FOV, while latter applied stable visual stimuli to users as rest frames. Future work should investigate whether adding dynamic pixel-level GRFs in provocative regions of the image is sufficient for VIMS mitigation. In that case, GRFs would be directly rendered on the image-plane pixels causing the most optical flow, thus interfering with the experience only where strictly necessary to offer stable visual cues.

Most of the computational cost in our method is from optical flow calculation in a pre-processing step. This can be largely reduced using more advanced cloud computing resources. Another challenge is the varied viewport dimensions among different headsets, which requires the pre-processing procedure to calculate the optical flow for different HMDs or with more fine-grained viewports that could later be combined based on the real viewport of the headset. Both solutions need extra storage and computation resources. However, the extra cost, as a trade-off, could save valuable and limited computation and storage resources in standalone HMDs and personal computers. Once sufficiently lightweight pre-trained models are developed and made available, they can be implemented directly into HMDs. This would enable true real-time optical flow estimation without the need for pre-processing.

\section{Conclusion} 
\label{sec:conclusion}

This work describes a novel method for estimating the optical flow experienced by a user in VR during the experience of a 360-degree video. This method involves a pre-processing step and allows for real-time estimation of optical flow at the viewport level for any user experiencing a pre-processed 360-degree video. Our method improves on previous work by offering customized visual flow estimation based on the user's current viewport, rather than using a single optical flow value for the entire equirectangular frame.

We conducted a small pilot study to verify the effectiveness of our method in reducing VIMS by presenting GRFs when high optical flow was detected. While the data from the pilot study is promising, further research is needed to definitively establish the relationship between optical flow and VIMS, as well as the effectiveness of GRFs in mitigating VIMS.

We conclude that our proposed method for estimating the optical flow in 360-degree videos offers a promising avenue for future work on improving user comfort and reducing simulator sickness in VR experiences.

\begin{acks}
This work was performed under award \#60NANB18D151N from the U.S. Department of Commerce, National Institute of Standards and Technology, Public Safety Communications Research Division.
\end{acks}

\bibliographystyle{ACM-Reference-Format}
\bibliography{bibliography}

\end{document}